\documentclass{aa}  

\usepackage{graphicx}
\usepackage{graphicx}
\usepackage{lscape}
\usepackage{longtable}
\usepackage{natbib}
\usepackage{color}
\usepackage{array} 
\bibpunct{(}{)}{;}{a}{}{,} 

\usepackage{txfonts}
\begin{document}

   \title{Kepler-447b: a hot-Jupiter with an extremely grazing transit\thanks{Based on observations collected at the German-Spanish Astronomical Center, Calar Alto, jointly operated by the Max- Planck-Institut fur Astronomie (Heidelberg) and the Instituto de Astrof\'isica de Andaluc\'ia (IAA-CSIC, Granada). } }


   \author{J. Lillo-Box\inst{1}, D. Barrado\inst{1}, N. C. Santos\inst{2,3,4}, L. Mancini\inst{5}, P. Figueira\inst{2,3}, S. Ciceri\inst{5}, Th. Henning\inst{5}
          }

  \institute{Depto. de Astrof\'isica, Centro de Astrobiolog\'ia (CSIC-INTA), ESAC campus 28691 Villanueva de la Ca\~nada (Madrid), Spain\\
              \email{Jorge.Lillo@cab.inta-csic.es}\and
Centro de Astrof\'{i}sica, Universidade do Porto, Rua das Estrelas, 4150-762 Porto, Portugal\and
Instituto de Astrof\' isica e Ci\^encias do Espa\c{c}o, Universidade do Porto, CAUP, Rua das Estrelas, PT4150-762 Porto, Portugal\and
Departamento de F\'{i}sica e Astronomia, Faculdade de Ci\^{e}ncias, Universidade do Porto, Portugal  \and
Max Planck Institute for Astronomy, K\"onigstuhl 17, 69117 Heidelberg, Germany  
            }
            
  \titlerunning{Kepler-447b: a hot-Jupiter planet with an extremely grazing transit}
\authorrunning{Lillo-Box et al.}
   \date{Accepted for publication in A\&A on February 10th, 2015.}

 
  \abstract
   { We present the radial velocity confirmation of the extrasolar planet Kepler-447b, initially detected as a candidate by the {\it Kepler} mission. In this work, we analyze { its transit signal} and the radial velocity data obtained with the Calar Alto Fiber-fed Echelle spectrograph (CAFE).  By simultaneously modeling { both datasets}, we obtain the orbital and physical properties of the system. According to our results, Kepler-447b is a Jupiter-mass planet ($M_p=1.37^{+0.48}_{-0.46}~M_{\rm Jup}$), with an estimated radius of $R_p=1.65^{+0.59}_{-0.56}~R_{\rm Jup}$ {(uncertainties provided in this work are $3\sigma$ unless specified)}. This translates into a { sub-Jupiter density}. The planet revolves every $\sim7.8$ days in a slightly eccentric orbit ($e=0.123^{+0.037}_{-0.036}$) around a G8V star with detected activity in the {\it Kepler} light curve. Kepler-447b transits its host with a large impact parameter ($b=1.076^{+0.112}_{-0.086}$), being one of the few planetary grazing transits confirmed so far and the first in the {\it Kepler} large crop of exoplanets. We estimate that only around $\sim$ 20\% of the projected planet disk occults the stellar disk. The relatively large uncertainties in the planet radius are due to the large impact parameter and short duration of the transit. { Planets with such an extremely large impact parameter can be used to detect and analyze interesting configurations such as additional perturbing bodies, stellar pulsations, rotation of a non-spherical planet, or polar spot-crossing events. All these scenarios would periodically modify the transit properties (depth, duration, and time of mid-transit), what could be detectable with sufficient accurate photometry. Short-cadence photometric data  (at the 1 minute level) would help in the search for these exotic configurations in grazing planetary transits like that of Kepler-447b.}
   }  


   \maketitle
%

\section{Introduction}

The unprecedented precision and long time span of the {\it Kepler} photometer have allowed the detection of exotic planetary systems such as very close-in hot-Jupiters around giant stars \citep[e.g. Kepler-91b, ][]{lillo-box14,lillo-box14c}, super-Earth size planets in the habitable zone of their parent stars \citep[e.g., Kepler-22b, ][]{borucki12}, terrestrial worlds \citep[e.g., Kepler-93b,][]{ballard14}, sub-Mercury size planets \citep[Kepler-37b, ][]{barclay13}, or very close-in packed planetary systems \citep[e.g., Kepler-11, ][]{lissauer11}. This, combined with the more than 150000 stars that were continuously observed, favors the detection and characterization of planets in unusual configurations, difficult or impossible to detect with other techniques.

In this context,  the {\it Kepler} light-curve of Kepler-447 (KOI-1800, KIC 11017901; RA=19$^h$01$^m$04$^s$.46, DEC=48$^{\circ}$33'36'') shows a V-shaped dip with a periodicity of $7.79430132 \pm 0.00000182$ days \citep{burke14}. According to \cite{huber14}, this object is a G8 Main-Sequence star slightly less massive  ($M_{\star}=0.764^{+0.145}_{-0.049}~M_{\odot}$) and smaller ($R_{\star}=0.872^{+0.49}_{-0.120}~R_{\odot}$) than the Sun. Usually, V-shaped eclipses are classified as false positives. They are mainly identified as eclipsing binaries with stars of similar sizes or grazing stellar eclipses. However, the combination of transit and radial velocity measurements can reveal the planetary nature of the transiting object. If a planetary mass is found for the transiting object, the V-shape is then explained as a grazing planetary eclipse. To date, just one planetary grazing eclipse has been reported and confirmed to accomplish the grazing criterion ($b+R_p/R_{\star}>1$) over $3\sigma$ \citep[WASP-67b,][]{hellier12,mancini14}. Also, few other planets are known to transit their star in near-grazing orbits, but with the grazing criterion accomplished below the $3\sigma$ level, as for instance WASP-34b \citep{smalley11} or HAT-P-27/WASP-40 \citep{beky11,anderson11}. So far, {\it Kepler} has discovered none.

The discovery of such grazing planetary transits opens an interesting window. In this configuration, any (even small) gravitational perturbation due to the presence of additional bodies in the system (like outer planets, exomoons, Trojans, etc.) would throw off the planet from its standard Keplerian orbit. Such effect would possibly induce a periodic variation in the impact parameter of the orbit that would be detectable in the transit data, so that they could become indirectly detectable \citep[e.g., ][]{kipping10}. The observational imprints of other effects (such as planet rotation and oblateness or stellar pulsations) could be maximized in such interesting configurations.

The present paper is part of an exoplanet survey carried out with instrumentation from the Calar Alto Observatory (CAHA). The follow-up of the {\it Kepler} candidates has already provided confirmation for two extrasolar planets by using the CAFE instrument, namely Kepler-91b \citep{lillo-box14c} and Kepler-432b \citep{ciceri14}\footnote{Also reported by \cite{ortiz14}}; has identified and analyzed some false positives and fast rotators \citep{lillo-box15a}; and has detected the secondary eclipse of WASP-10b with OMEGA2000 \citep{cruz14}. The present work, regarding Kepler-447, is organized as follows. In Sect. \S~\ref{sec:observations} we summarize the ground- and space-based data used in this study. In Sect. \S~\ref{sec:analysis} we analyze the data by obtaining new spectrum-based stellar parameters (\S~\ref{sec:stellarparams}) and modeling the radial velocity  and transit signals in a simultaneous fit (\S~\ref{sec:joint}). { In Sect. \S~\ref{sec:properties} we discuss the properties of this planetary system. In Sect. \S~\ref{sec:configurations} we discuss the possible (and exotic) configurations detectable in these extremely grazing planetary transits and  review an important systematic effect causing the presence of outliers inside the region of short-duration transits.} We summarize the conclusions of our analysis in Sect. \S~\ref{sec:conclusions}.

   \begin{figure*}[htbf]
   \centering
   \includegraphics[width=1\textwidth]{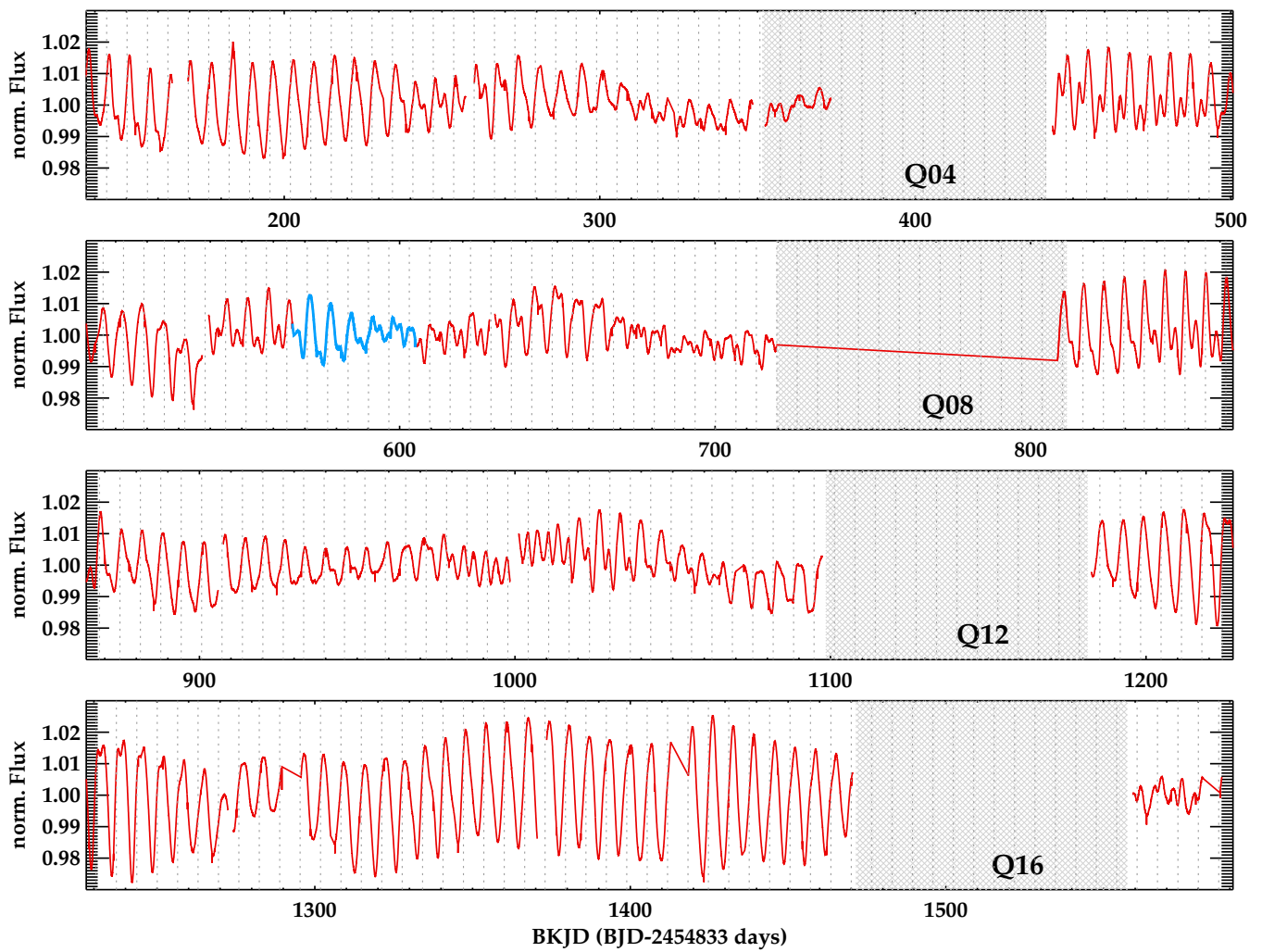}   
   \caption{Photometric time series of Kepler-447 as obtained by the {\it Kepler} telescope. The simple aperture photometry (SAP) is plotted in red. The quarters in which the source felt in the death detector channels of the CCD are marked as shaded regions (see Sect.~\ref{sec:kepler}). The vertical dotted lines are plotted with a periodicity equal to the measured rotational period by \cite{mcquillan13b}. The orbital periods 55-59 are marked as blue filled circles (see Sect.~\ref{sec:outliers}).}
   \label{fig:LCcomplete}
   \end{figure*}

\section{Observations and data reduction \label{sec:observations}}

\subsection{{\it Kepler} photometry\label{sec:kepler}}
We have retrieved the {\it Kepler} photometry of this target from quarters Q1-Q17 (more than 1400 days, only with the long-cadence mode, 29.4 minutes). The publicly available data were downloaded from the {\it Kepler} MAST\footnote{https://archive.stsci.edu/kepler/data\_search/search.php} (Milkulski Archive for Space Telescopes). { During Q4, the detector channels 5, 6, 7, and 8 in Module 3 of the charge-coupled device (CCD) failed. Consequently, any source falling in this detectors at any quarter could not be observed. Due to the rotation of the spacecraft, Kepler-447 felt in this broken down module on Q4, Q8, Q12, and Q16. Thus, no data were acquired for this KOI during those quarters.}

We used the { simple aperture photometry (SAP)} flux and its corresponding uncertainties provided by the {\it Kepler} team to compute the final light-curve (LC). {\rm We removed from this analysis any datapoint with a quality flag in the processed LC (SAP\_QUALITY) equal to 128 (cosmic ray correction). In Sect.~\S~\ref{sec:outliers} we analyze the source and the effect of this data points on the LC}. Artificial (well-known) large trends in the {\it Kepler} data { and the observed modulations due to stellar activity} were significant in this case. Consequently, we needed to detrend the fluxes to analyze the transit signal. We used a cubic spline function to model the out-of-transit modulations, selecting the nodes by measuring the mean fluxes of one-day bins. This simple and quick approach provided a clean and flat light-curve in the out-of-transit region while keeping the transit unperturbed. 

\subsection{High-spatial resolution imaging}

The host star candidate Kepler-447 was part of our high-resolution imaging survey of {\it Kepler} candidates \citep{lillo-box12,lillo-box14b}. We obtained lucky-imaging images\footnote{We used the AstraLux instrument at Calar Alto Observatory.} for more than 170 KOIs with enough magnitude contrasts and spatial resolution to statistically discard possible stellar configurations mimicking a planetary transit. 

We defined the blended source confidence parameter (BSC) as the probability for a given object observed with high-resolution techniques and without any detected companions, to be actually isolated \citep{lillo-box14b}. In the case of Kepler-447, the high-resolution images and their posterior analysis revealed a BSC value of 99.9\%. This implies a 0.1\% of probability of having a blended (undetected) source contaminating the {\it Kepler} light-curve. Thus, we will assume along this paper that Kepler-447 is an isolated target.

\subsection{High-resolution spectroscopy}

This star was observed with the CAFE instrument \citep{aceituno13} installed at the 2.2m telescope in Calar Alto Observatory (Almer\'ia, Spain) as part of our follow-up of {\it Kepler} candidates\footnote{This follow-up has already confirmed the planetary nature of Kepler-91~b \citep{lillo-box14,lillo-box14c} and Kepler-432~b \citep{ciceri14}.}. This instrument has no movable pieces that could introduce artificial signals in the radial velocity. Also, it is located in an isolated chamber and its spectral coverage is fixed in the range $\sim$ 4000-9500 \AA, with a spectral resolution in the range $R=[59000,67000]$, depending on the order. The chamber is monitored in temperature, pressure, and humidity to check for possible relevant changes during the night. 

 We obtained 21 high-resolution spectra of the target, sampling the different orbital phases of the planet candidate. The spectra were reduced by using the dedicated pipeline  provided by the observatory\footnote{This pipeline was developed by Sebasti\'an S\'anchez.}, explained in \cite{aceituno13}. We did not performed any dark correction and we used thorium-argon (ThAr) spectra obtained after every science spectrum to wavelength-calibrate the data.
 
 We cross-correlated the observed spectra against a weighted binary mask to obtain the radial velocity (RV) of each epoch \citep{baranne96}. This mask was designed by using a solar spectrum obtained from the solar atlas provided by BASS 2000\footnote{http://bass2000.obspm.fr/solar\_spect.php} \citep{delbouille72}. From this spectrum, we selected a series of isolated, sharp, and high-contrast absorption lines. In total, our mask includes around 2100 lines in the CAFE spectral range (4000-9500 \AA ). We used a velocity range of $\pm30$ km/s around the expected radial velocity value for each case. The radial velocity of each spectrum is measured as the center of a gaussian fit to the resulting cross-correlation function (CCF), sum of all CCFs of the different orders. We corrected this velocity from the barycentric Earth radial velocity (BERV) obtained from the corresponding Julian date at mid-exposure. In Table~\ref{tab:rvobs}, we show the observing data and the final RV values for each epoch.

\section{Analysis \label{sec:analysis}}

\subsection{Stellar properties: rotation and physical parameters \label{sec:stellarparams}}

\begin{figure}[htbf]
   \centering
   \includegraphics[width=0.5\textwidth]{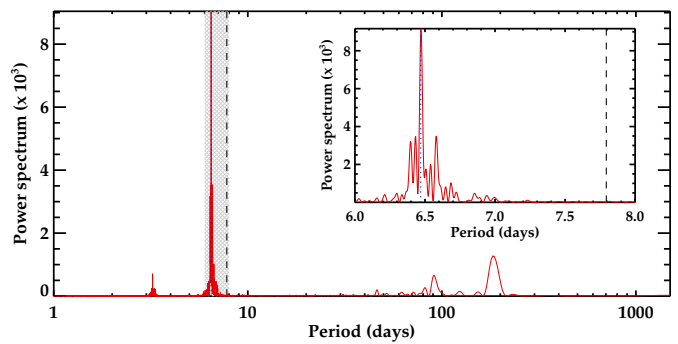}   
   \caption{Lomb-Scargle periodogram of the complete {\it Kepler} light curve (red line). A zoom into the shaded region around the rotational period of the star ($\sim$6.5) days is shown in the small panel. The vertical dashed line corresponds to the orbital period, while the dotted vertical line is the measured rotational period by \cite{mcquillan13b}.}
   \label{fig:LCperiodogram}
   \end{figure}

{ The SAP flux extracted by the {\it Kepler} pipeline shows a clear variability at the level of $\sim$ 4\% (peak to peak, see Fig.~\ref{fig:LCcomplete}). These type of variations have been detected in many other sources with well-sampled photometric time series.  They are related to the rotation of the star and the presence of stellar spots on its surface.  In particular, \cite{mcquillan13b} analyzed the photometric modulation caused by starspots in the {\it Kepler} sample of planet candidates. They calculated the rotational periods for several hundreds of KOIs by using the autocorrelation function technique (ACF), described in \cite{mcquillan13a}. In particular, they obtained a rotational period for Kepler-447 of $P_{\rm rot} = 6.459 \pm 0.003$ days.  Added to this, we have computed the Lomb-Scargle periodogram of the light curve (see Fig.~\ref{fig:LCperiodogram}). It shows a clear set of peaks around $P_{\rm rot}$, with the highest one at $P_{\rm peak}=6.4723\pm0.0003$~days. Additional but less significant peaks are found at larger periodicities, but their analysis is out of the scope of the current work. Both the orbital and the stellar rotation share similar periodicities. This could be due to some on-going synchronization between the stellar rotation and the orbital period due to similar processes studied in binary stars, such as tidal friction \citep{zahn66} or hydrodynamical mechanisms \citep{tassoul87}. 

At this point, it is worth to mention that the presence of stellar spots producing the high-amplitude photometric modulations implies the existence of stellar activity. This could contaminate the transit signal with possible spot crossing events such as those detected in other hot Jupiters around active stars \citep[e.g.,][]{silva-valio11,desert11b}. These signals can also be used to measure sizes, temperatures, and positions of the spots \citep{silva03}, as well as the spin-orbit alignment between the planetary orbit and the stellar rotation axis \citep[e.g., ][]{sanchis-ojeda11}. However, the long-cadence data obtained for Kepler-447 and the short duration of its transit prevents this kind of studies in this system with the current data. 
}

{ Added to the stellar rotation, we used the CAFE spectra to obtain the physical properties of the host star.} We combined all high-resolution spectra by shifting them according to their measured RV to obtain a moderate signal-to-noise (S/N) spectrum. The final combination provides S/N $= 65$. This moderate S/N spectrum was used to obtain the stellar parameters of the host star. We derived the effective temperature, surface gravity, and metallicity by using spectral synthesis, using the Fe {\sc i} spectral lines as described in \cite{pepe11}. The results show that Kepler-447 is a G8V star with $T_{\rm eff}=5493\pm62$~K, $\log{g}=4.40\pm0.10$, and [Fe/H] $=+0.07\pm0.05$. These values, agree with the photometrically obtained by \cite{huber14} except for the metallicity, being sub-Solar in the latter work but with much larger uncertainties ([Fe/H]$_{\rm H14}=-0.40^{+0.36}_{-0.26}$). In Table~\ref{tab:stellarparams}, we show both sets of stellar properties.

By using these parameters, we can estimate the stellar mass and radius by applying the parametrization presented in \cite{torres10}. We estimated the uncertainties in these parameters by running $10^5$ Monte Carlo trials obtained from a gaussian distribution of the input values (i.e., $T_{\rm eff}$, $\log{g}$, and [Fe/H]) and the coefficients of the parametrization \citep[see Table 1 in][]{torres10}. We then take the 68\% limits ($1\sigma$) as the estimated uncertainties. According to this, we obtain $M_{\star} = 1.00\pm0.21~M_{\odot}$ and $R_{\star} = 1.03\pm0.16~R_{\odot}$. These values will be assumed along this work for further estimation of orbital and physical properties of the transiting companion. For completeness, we will also derive those properties by assuming the values provided by \cite{huber14}.

\subsection{Post-processing of the radial velocity and transit data \label{sec:considerations}}
CAFE data were collected during several runs in 2011, 2012, and 2013. Due to the long time span, we must account for several changes in the ThAr lamps along the whole set of observations when comparing all RV epochs. In total, we can divide our data into three groups corresponding to three different ThAr lamps. We can correct the effect of the different lamps by applying RV offsets  between the samples. In total, we would need two offsets  ($v_{\rm off1}$ and $v_{\rm off2}$) to set all three RV sets at the same level. These offsets will be included as free parameters in the RV fit, explained in the next section.

By phase folding the detrended light-curve with the detected periodicity, we found { few data points clearly over the transit signal. The source of these outliers (once the flagged data points with cosmic rays correction have been removed) is likely due to other instrumental effects. We ignored these flagged points when fitting the transit signal.} Another remarkable feature to point out is that the transit is short ($\sim 1.13$ hours) and V-shaped, rather than U-shaped as expected for planets. This suggests a large impact parameter, with the transit being grazing.

  \begin{table*}[htbf]
\setlength{\extrarowheight}{2pt}
\scriptsize
\caption{Measured radial velocities with CAFE.}
\label{tab:rvobs}
\centering 
\begin{tabular}{cccc|cccc|cccc}     
\hline\hline
 Julian Date & $\overline{\rm S/N}$ & Phase & RV & Julian Date & $\overline{\rm S/N}$ & Phase & RV & Julian Date & $\overline{\rm S/N}$ & Phase & RV       \\ 
(days)-2456000 & & & (km/s) & (days)-2456000 & & & (km/s) & (days)-2456000 & & & (km/s) \\ \hline
           76.593035  &       26.0  &           0.941  &              $1.421^{+0.027}_{-0.027}$ &           90.523633  &       23.6  &           0.729  &              $1.545^{+0.032}_{-0.032}$ &          812.556916  &       11.9  &           0.365  &                $1.209^{+0.042}_{-0.042}$ \\
           77.536041  &       23.5  &           0.062  &              $1.335^{+0.032}_{-0.032}$ &           91.376159  &       18.8  &           0.838  &              $1.558^{+0.035}_{-0.035}$ &          817.599630  &        7.0  &           0.012  &                $1.301^{+0.056}_{-0.056}$ \\
           80.543287  &       25.7  &           0.448  &              $1.361^{+0.031}_{-0.031}$ &          523.571588  &       11.6  &           0.288  &              $0.928^{+0.043}_{-0.043}$ &          818.553484  &       10.9  &           0.134  &                $1.188^{+0.034}_{-0.034}$ \\
           88.515082  &       15.3  &           0.471  &              $1.392^{+0.031}_{-0.031}$ &          597.302622  &       18.2  &           0.748  &              $1.136^{+0.048}_{-0.048}$ &          818.585851  &       11.4  &           0.138  &                $1.168^{+0.030}_{-0.030}$ \\
           88.549981  &       17.4  &           0.475  &              $1.389^{+0.033}_{-0.033}$ &          801.517625  &        9.5  &           0.948  &              $1.308^{+0.036}_{-0.036}$ &          823.582348  &       13.8  &           0.778  &                $1.462^{+0.025}_{-0.025}$ \\
           89.381498  &       17.1  &           0.582  &              $1.469^{+0.029}_{-0.029}$ &          801.552053  &       10.9  &           0.953  &              $1.396^{+0.045}_{-0.045}$ &          842.549205  &       11.5  &           0.213  &                $1.152^{+0.037}_{-0.037}$ \\
           89.416506  &       13.8  &           0.587  &              $1.411^{+0.040}_{-0.040}$ &          812.524544  &       11.9  &           0.360  &              $1.303^{+0.033}_{-0.033}$ &          859.401764  &        7.3  &           0.375  &                $1.188^{+0.039}_{-0.039}$ \\
\hline 
\end{tabular}
\end{table*}

\subsection{Origin of the radial velocity signal}

\begin{figure*}[hbft]
   \centering
   \includegraphics[width=0.5\textwidth]{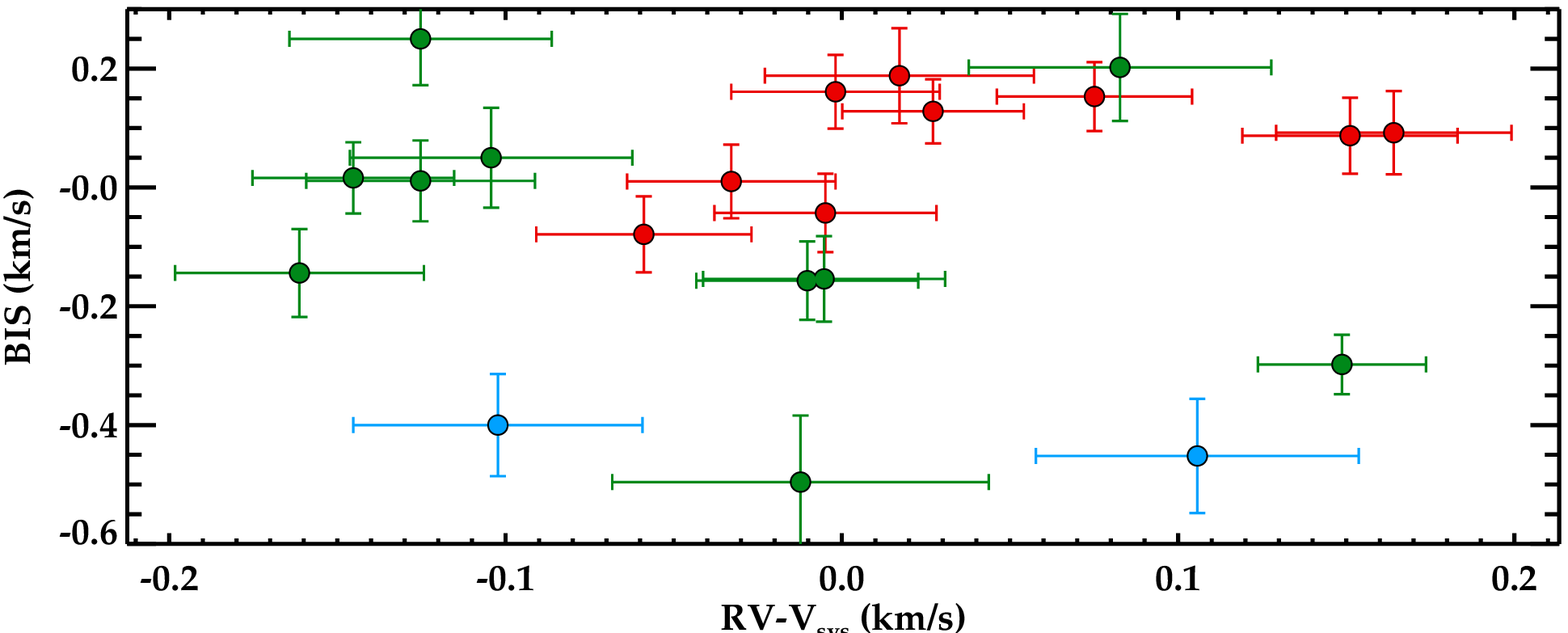}
   \includegraphics[width=0.5\textwidth]{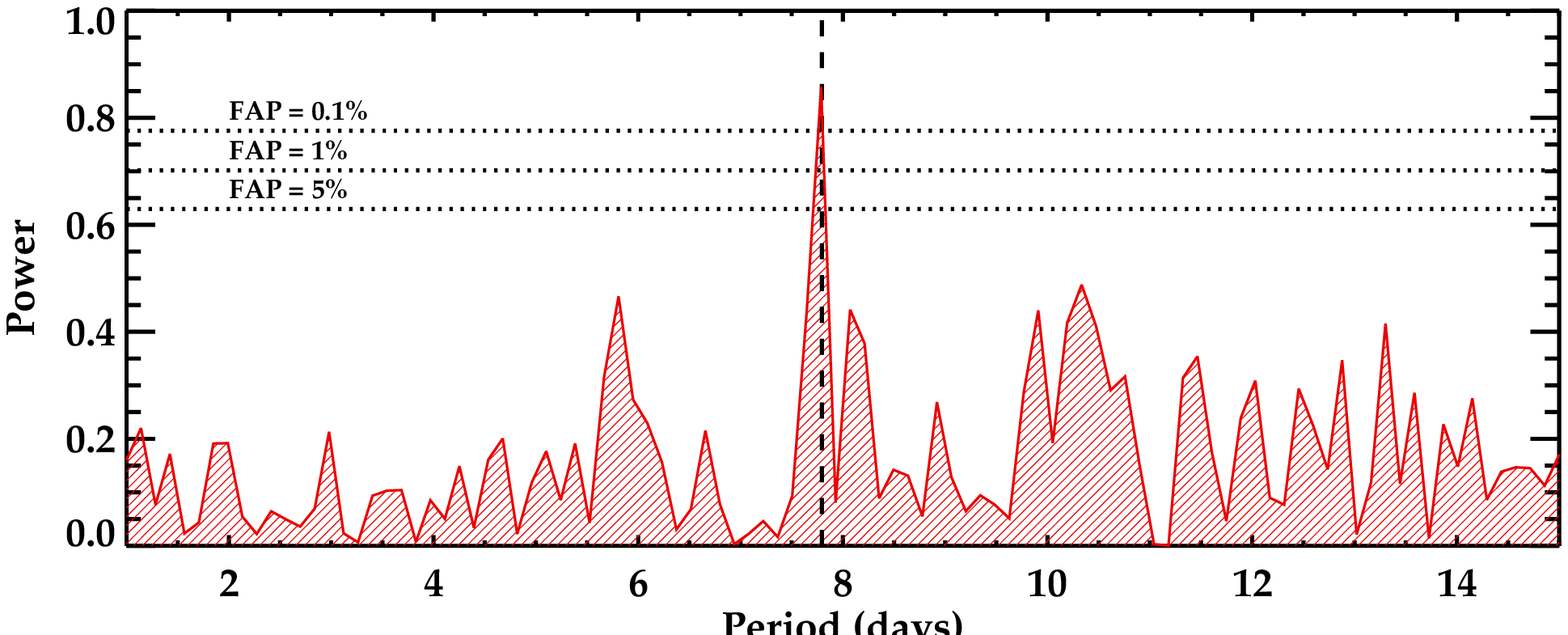}
   \caption{Upper panel: bisector analysis as a function of the measured radial velocity obtained during runs on 2012 (red symbols), 2013 (light blue), and 2014 (green). Lower panel: periodogram of the radial velocity data showing the significant peak at the corresponding transit period (vertical dashed line).}
   \label{fig:bis}
   \end{figure*}

We applied several line profile tests\footnote{By using the python code provided by P. Figueira (available at https://bitbucket.org/pedrofigueira/line-profile-indicators), explained in \cite{figueira13} and presented in \cite{santos14}.} to check for possible dependencies of the RV variation with stellar activity. We found no positive correlations between both sets of values in any of the performed tests. In particular, the bisector analysis \citep[BIS, ][]{queloz01} provides a Pearson's correlation coefficient of 0.007, thus showing no correlation of the RV with the line profile indicator. We show the BIS values against their corresponding RVs in Fig.~\ref{fig:bis}. This may indicate that the RV is not correlated with possible line profile asymmetries caused by stellar inhomogeneities such as spots.

We have also computed the periodogram of the radial velocity data (see lower panel of Fig.~\ref{fig:bis}). It shows a relevant peak at the expected period from the transit analysis, with a false-alarm probability (FAP) of ${\rm FAP}<0.01$ \%. This also indicates that the object transiting the host star is actually producing the RV variations.

\subsection{Simultaneous fit of the radial velocity and transit signals \label{sec:joint}}

Due to the aforementioned peculiar characteristics of this system, we decided to perform a simultaneous fit of the RV and transit data. This is because both effects share key parameters that could importantly affect the results of the other dataset (in particular the eccentricity and the argument of the periastron). The orbital period ($P_{\rm orb}$) and the mid-transit time ($T_0$) were fixed to the values measured by \cite{burke14}. 

In total, ten free parameters are needed to model both datasets: radial velocity semi-amplitude ($K$), eccentricity ($e$), argument of periastron ($\omega$), systemic velocity of the system ($V_{\rm sys}$)\footnote{We note that this $V_{\rm sys}$ is not the absolute systemic velocity of the system since it is not corrected from the possible instrumental offsets, although this should be of some tens of m/s.}, two RV offsets ($v_{\rm off1}$ and $v_{\rm off2}$, see Sect.~\S~\ref{sec:considerations}), semi-major axis to stellar radius ($a/R_{\star}$), planet-to-star radius ratio ($R_p/R_{\star}$), inclination ($i$), and a phase offset in the transit center to include possible unaccounted uncertainties in the determination of the transit epoch ($\phi_{\rm off}$). 

Among these free parameters, six are needed to model the transit signal by using the \cite{mandel02} formulation (i.e., $R_p/R_{\star}$, $e$, $\omega$, $a/R_{\star}$, $i$, and $\phi_{\rm off}$). { According to the recommendations from \cite{muller13} for high impact parameter transits, we fixed the limb-darkening coefficients to the theoretical values. }We performed a trilinear interpolation of the determined stellar properties ($T_{\rm eff}$, $\log{g}$, and [Fe/H]) on the { non-linear four terms} limb-darkening values calculated by \cite{claret11} for the {\it Kepler} band. { The resulting values used in the transit fit are $a_1=0.5895$, $a_2=-0.2477$, $a_3=0.8538$, and $a_4=-0.4166$.}

Due to the short duration of the transit ($\sim$ 1.13 hour) and the long cadence data obtained for this object (i.e., 29.4 minutes), there is a remarkable effect to take into account in the transit modeling. {It is based in the fact that we try to fit long-cadence (timely binned) data with a theoretical transit model with infinite resolution (i.e., unbinned). In practice, the long-cadence data smooths the transit signal, producing a broader transit. If the transit is grazing, this also translates into a shallower depth. This effect was already highlighted by \cite{kipping10}, who warned about large errors in the derived parameters if unaccounted. By definition, this binning effect is larger for short (with sampling step of the order of the transit duration) and grazing transits, such as the present case. As explained by \cite{kipping10}, there are several ways to correct for it. The simplest (but most efficient)  is to compute the light-curve model with a short cadence sampling (e.g., 1 minute) and then bin the model with the actual cadence of the observed data. We used this approach in our analysis.}

Regarding the RV, we used the well-known model for the RV variations of a star surrounded by a single planet
\begin{equation}
	V(t) = V_{\rm sys} + K [ \cos{(\nu(t) + \omega) + e\cos{\omega}}],
\end{equation}
\noindent where $\nu(t)$ is the true anomaly of the planet. The true anomaly at each epoch was obtained by solving the Kepler equation according to the eccentricity and phase of the planet. 

We used our genetic algorithm {\it GAbox}\footnote{{\it GAbox} is a fitting tool for any kind of scientific modeling and has already been used as spectral energy distribution fitter \citep{riviere-marichalar13, riviere-marichalar14}, transit and light-curve modulations fitting \citep{lillo-box14}, and to extract the RV of high-resolution spectra \citep{lillo-box14c}.} to explore the parameter space and find the set of parameters that best reproduces the current data. {\it GAbox} allows for a clever exploration of this high-dimensional problem. We broadly restricted the parameter space according to our knowledge of the data. The adopted ranges for each of the free parameters are summarized in Table~\ref{tab:results}. 

\begin{figure*}[htbf]
   \centering
  \includegraphics[width=\textwidth]{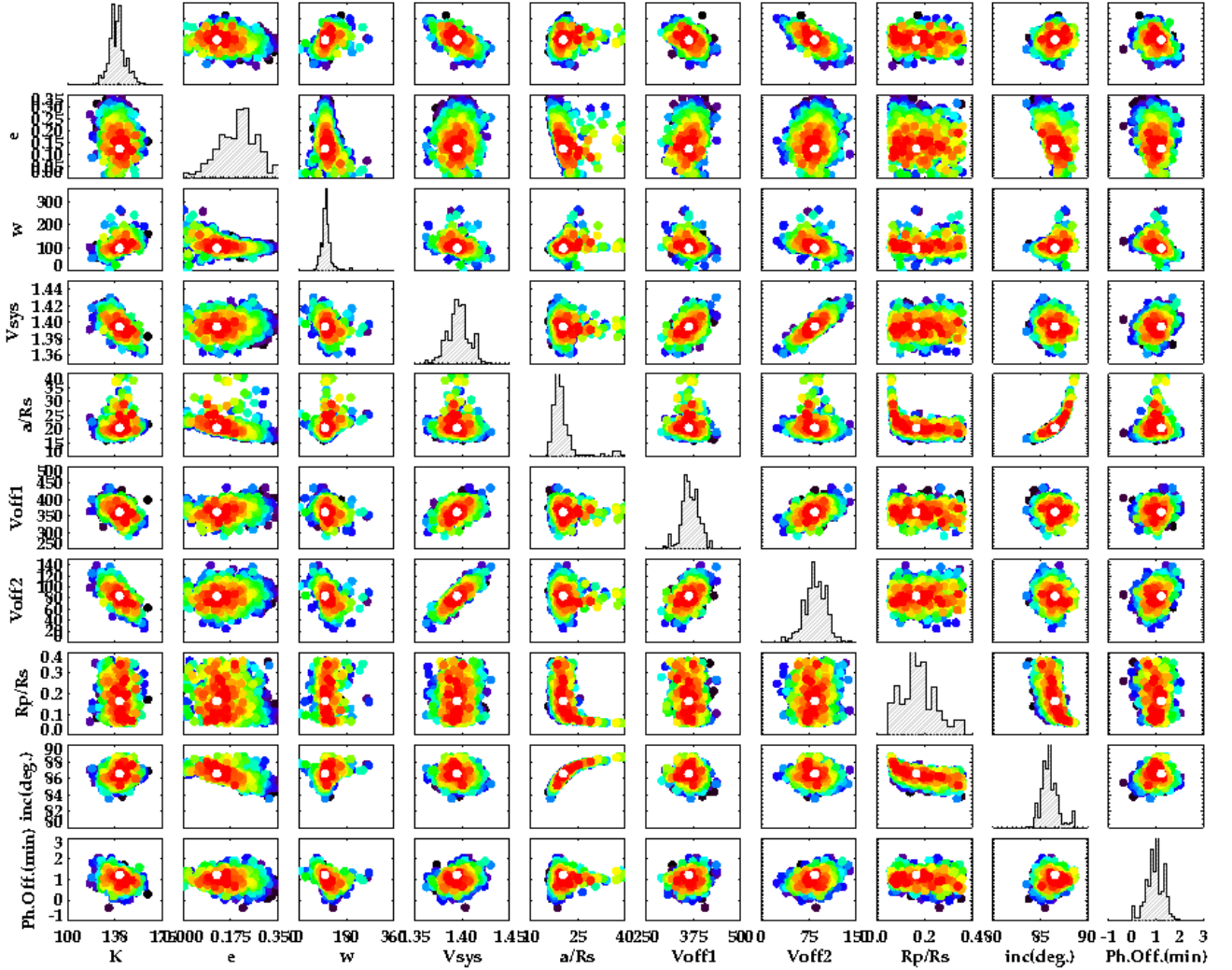}
   \caption{Parameter-parameter representation of the 1500 convergence solutions of the transit fitting using {\it GAbox}. In color-code we represent the combined reduced chi-square statistic of the RV and transit fitting models (dark colors represent poorer chi-square models). The diagonal panels show the histograms of the distribution of the individual parameters within all solutions. The white circle represents the least-square model.}
   \label{fig:gabox}
   \end{figure*}

Due to the different number of data points on each data set ($N_{\rm RV}=21$ and $N_{\rm LC}=994$), we used the reduced chi-square as the minimization parameter in order to equally weight both effects. Mathematically, this minimization parameter is expressed as $\chi^2_{\rm red}=\chi^2_{\rm red,RV} + \chi^2_{\rm red,LC}$. The distribution of the 1500 convergence solutions  found by {\it GAbox}in the parameter-parameter space is plotted in Fig.~\ref{fig:gabox}. In this figure we can see that all parameters but the planet radius are very well constrained. However, as it is shown by the color-code in Fig.~\ref{fig:gabox}, the least-square solution is located in the middle of the preferred region, suggesting that the adopted solution is a good compromise for all parameters ($\chi^2_{\rm red,RV}=1.2$ and $\chi^2_{\rm red,LC}=1.8$). We adopted this least-square set of parameters as our final solution. 

\begin{figure*}[htbf]
   \centering
   \includegraphics[width=0.49\textwidth]{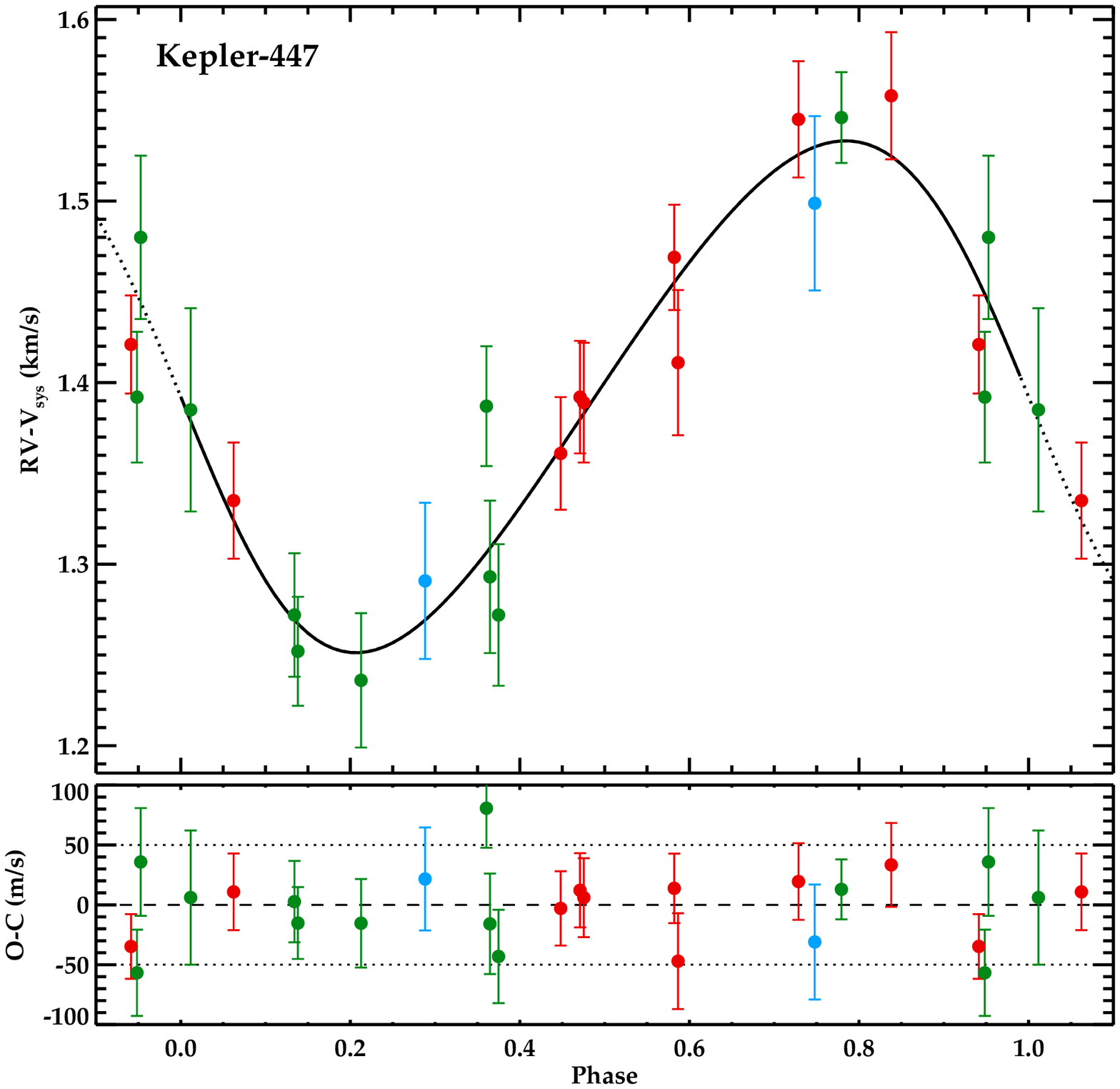}   \includegraphics[width=0.49\textwidth]{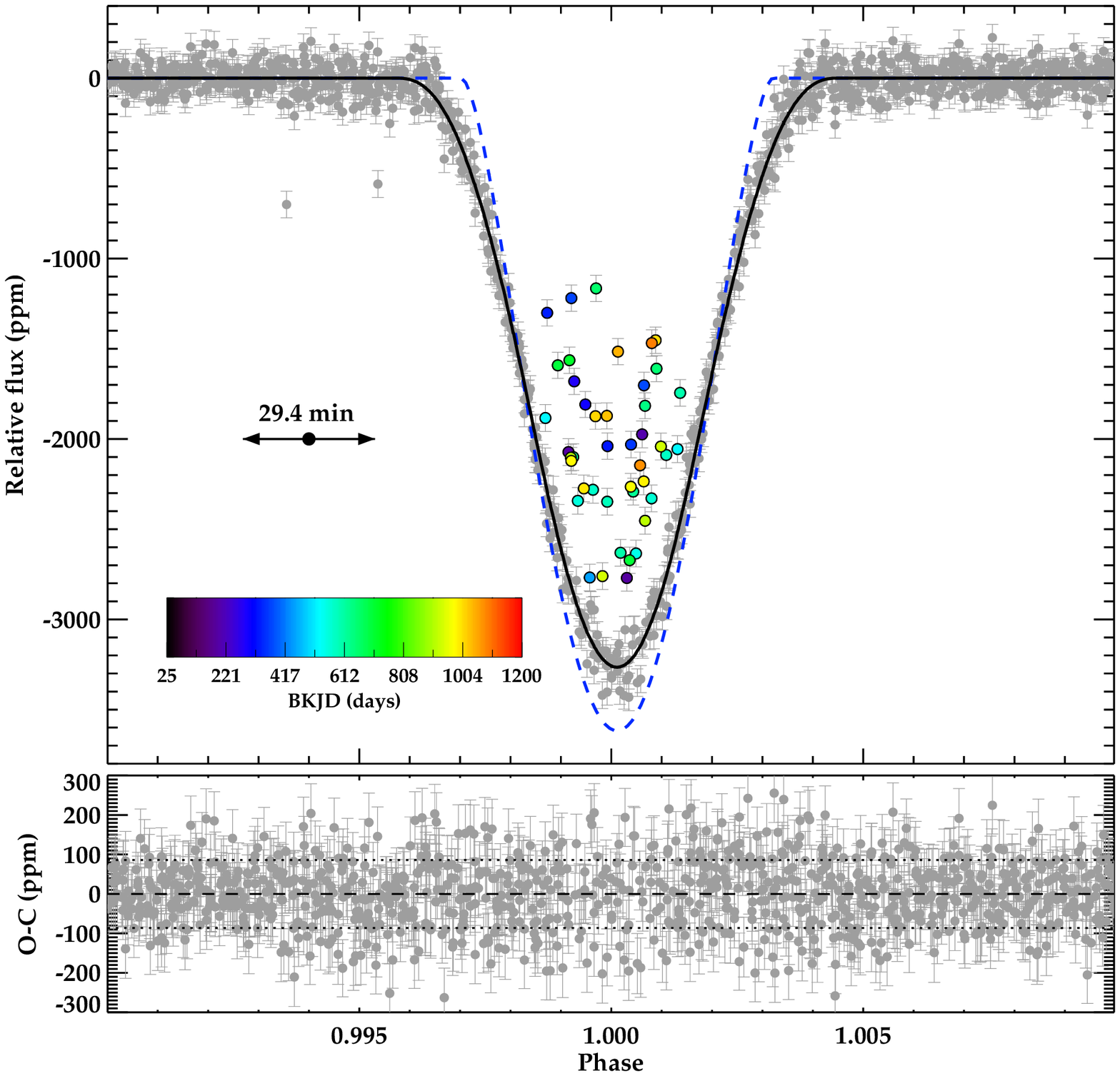}
   \caption{\textbf{Left panel:} Phase-folded radial velocity data obtained during runs on 2012 (red symbols), 2013 (light blue), and 2014 (green). The black line shows the best fit model. The lower left panel shows the residuals of the model fit, having a rms of 31 m/s. \textbf{Right panel:} transit fitting of Kepler-447b. In the upper panel, we show the detected transit in the {\it Kepler} light-curve. We mark in color the outliers of the transit that have been removed from the fitting process (see Sect.~\ref{sec:considerations}), { and that are mostly due to the misidentification of cosmic rays by the \emph{Kepler} pipeline}. The color-code is shown in the color bar and relates to the corresponding Barycentric {\it Kepler} Julian Date (BKJD $ = $ BJD-2454833 days). The final fitted model is represented by the solid black line. We have also included with a dashed blue line the original (not binned) model from which it is calculated (see Sect.~\ref{sec:joint}). The bottom panel shows the residuals of the fit, with a rms of 101 ppm.}
   \label{fig:rv}
   \end{figure*}

The uncertainties were obtained by running Monte-Carlo Markov Chain (MCMC) simulations with the Metropolis-Hasting algorithm. We ran $10^6$ steps and discarded the 10\% firsts to avoid dependence on the priors, which were set to the {\it GAbox} solution. We used $3\sigma$ levels as the adopted uncertainties. The final parameters are shown in Table~\ref{tab:results}, and the results of this fit are presented in Fig.~\ref{fig:rv}. 

Given the obtained parameters in Table~\ref{tab:results}, we can derive the mass of the companion revolving around Kepler-447 by using its relation with the semi-amplitude of the RV signal
\begin{equation}
K^3 = \frac{2\pi G}{P(1-e^2)^{3/2}} \left[ \frac{M_p^3\sin^3{i}}{(M_p+M_{\star})^2}\right],
\end{equation}

\noindent where the value inside the brackets is the so-called mass function ($F_M$). By assuming the stellar properties, we can obtain absolute values for the planetary mass and radius. Due to the relative disagreement between the stellar mass and radius obtained by \cite{huber14} and the present work (although still within the $1\sigma$ uncertainties), we decided to get absolute parameters for both sets of stellar properties. However, we will refer to the final results as those obtained from our spectroscopic values (see Sect.~\S~\ref{sec:stellarparams}). We thus obtain a  mass of $M_p = 1.35^{+0.48}_{-0.46}~M_{\rm Jup}$ for the companion object, establishing its planetary nature. The large uncertainty is consequence of the uncertainty in the stellar mass. Similarly, by using  the $R_p/R_{\star}$ ratio, we obtain a planetary radius of  $R_p=1.65^{+0.59}_{-0.56}~R_{\rm Jup}$. We should note that $R_p/R_{\star}$ is not very well constrained due to the short duration and grazing shape of the transit signal (see Fig.~\ref{fig:gabox}). This is translated in a relatively large uncertainty for this parameter.

\section{Discussion \label{sec:discussion}}

\subsection{Derived properties of the planetary system \label{sec:properties}}

We have shown that the companion transiting Kepler-447 has a planetary mass of $M_p = 1.37^{+0.48}_{-0.46}~M_{\rm Jup}$ and a large radius of $R_p=1.65^{+0.59}_{-0.56}~R_{\rm Jup}$. According to these values, the resulting mean density of the planet would be $\rho_p=0.30^{+0.71}_{-0.24}~\rho_{\rm Jup}$. This low density could indicate an inflated atmosphere for the planet. Although several inflation mechanisms could be taking place, the eccentricity found for this close-in giant planet suggests that tidal heating could be the main mechanism \citep[e.g., ][]{jackson08}. Note that at periastron passage, the planet approaches its host star at a distance of $r_{\rm per}/R_{\star}=17.89^{+0.80}_{-0.75}$. The non-negligible eccentricity found could be due to a third body in the system, preventing the circularization process and enhancing the tidal heating. 

\begin{figure*}[htbf]
   \centering
   \includegraphics[width=0.5\textwidth]{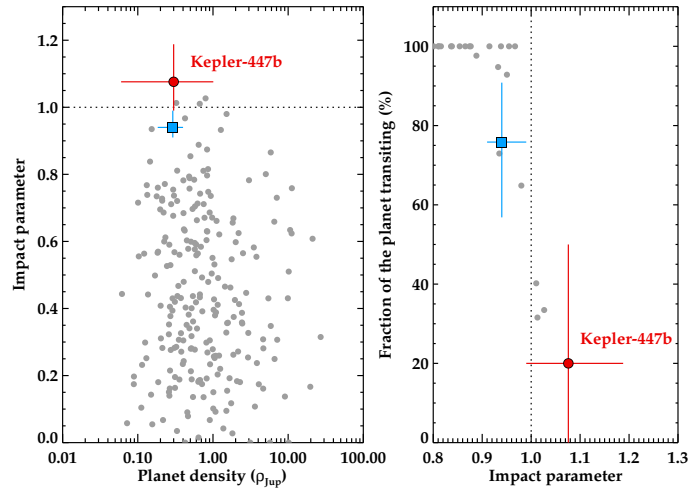}
   \caption{\textbf{Left panel:} impact parameter versus planetary density of known exoplanets with available information. \textbf{Right panel:} fraction of the projected planet transiting its host star as a function of the impact parameter. In both panels, Kepler-447b has been highlighted in red color and big circle symbol, including uncertainties. For reference, WASP-67b is also highlighted in light blue color with a square symbol. For clarity reasons, we do not include uncertainties for the other planets.}
   \label{fig:impact}
   \end{figure*}

The semi-major axis of the planetary orbit is  $a =0.069^{+0.006}_{-0.008}$ AU as calculated by using the third Kepler law (assuming the mass of the host and the orbital period). We then obtain an impact parameter of $b = 1.076^{+0.112}_{-0.086}$, confirming that the transit is highly grazing.  At mid-transit, only around $20$\% of the planet disk actually eclipses the star (i.e., $A_{\rm ecl}/A_{\rm pl}=0.20^{+0.30}_{-0.23}$). This explains the V-shape of the transit for this planetary mass companion. Given this impact parameter and the calculated planet-to-star radius ratio, we obtain a grazing criterion of $b+R_p/R_{\star}=1.24^{+0.12}_{-0.10}$ ($3\sigma$ uncertainties), clearly above the $3\sigma$ level. This confirms that Kepler-447b is the second known planet with a grazing transit, and the first in the {\it Kepler} sample. The first planet known to accomplish this criterion was WASP-67b \citep{hellier12,mancini14} which was analyzed with short-cadence data from the ground. According to the Exoplanet Archive\footnote{http://exoplanet.eu}, Kepler-447b would be the most grazing transiting exoplanet found to date, having the least fraction of its projected area covering the stellar disk. We have illustrated this in Fig.~\ref{fig:impact}.

\begin{figure*}[htbf]
   \centering
   \includegraphics[width=0.5\textwidth]{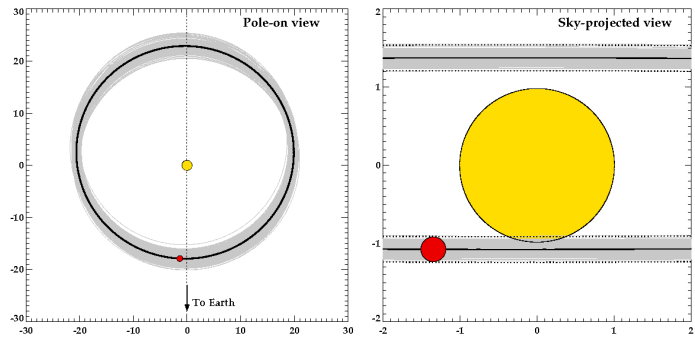}
   \caption{Orbital scheme the solution found for the orbit of the planet Kepler-447b (red filled circle). {\it Left panel}: Pole-on view of the orbit (separation from the star in units of the stellar radius). We show 500 orbits by bootstrapping the parameters inside their uncertainty limits.  {\it Right panel}: Face on view of the orbit as projected in the sky (separation from the star in units of the stellar radius). We have marked the path of the planet between the dotted lines. The gray solid lines represent 500 orbits by bootstrapping the impact parameter. We can see that the transit is always grazing independently of the parameters taken inside the uncertainty limits.}
   \label{fig:orbits}
   \end{figure*}

In Fig.~\ref{fig:orbits}, we show a scheme of the pole-on and face-on views of the system. As it is shown, the planet does not produce a secondary eclipse due to the slightly inclined and the non-circular orbit. This is in agreement with the lack of a secondary eclipse for such a large and close-in planet.

In Table~\ref{tab:derived} we present all derived physical and orbital properties for this system by assuming both sets of stellar parameters from our own analysis of the high-resolution and moderate S/N spectrum of the star (solution A) and from \cite{huber14}, solution B.

\subsection{Relevant configurations detectable in extremely grazing planetary transits \label{sec:configurations}}

{ The extremely large impact parameter that we have derived for the transit of the planet Kepler-447b means that interesting effects non detectable in transiting planets with shorter impact parameters can show up. In this section we summarize the most interesting effects that could be detectable in this kind of architectures:}

\begin{enumerate}

\item \textit{Additional bodies}.- Due to the extremely high impact parameter of the orbit, small gravitational perturbations provoked by additional bodies in the system could induce the (projected) libration of the planet. { This would induce periodic variations in the depth, duration, or transit time, depending on the relative position of both bodies. Among the different configurations with an additional third body in the system, the most common configuration could be} the existence of an \textit{outer (non-transiting) planet} in the system. In this case, we should also detect transit timing and depth variations, depending on the architecture of the system. However, { short-cadence data is needed to analyze each transit individually and detect these variations.} Another hypothesis can be the presence of a natural satellite orbiting around the planet (i.e., an \textit{exomoon}) that would induce an orbital wobble of the planet \citep{kipping09b} and thus the variation of the impact parameter on every orbit. { In the most extreme case, the transit would not occur in some planetary revolutions due to this wobble of the orbital plane}. Also, massive \textit{Trojan} bodies can perturb the orbit of the planet as it has been investigated, for instance, by \cite{laughlin02}.  \\

\item {\it Stellar pulsation}.- If the star is pulsating, the stellar radius will slightly increase and decrease, changing the $R_p/R_{\star}$ ratio and thus the fraction of the stellar disk occulted by the planet (or equivalently changing its impact parameter). { These stellar pulsations should also be detectable in the out-of-transit time interval as periodic modulations of the light curve.}  \\

\item {\it Planet rotation and oblateness}.- If the planet has a slightly ellipsoidal shape (i.e., non-negligible oblateness) and its self-rotation period is not coupled to the orbital  period, the projected planet area could differ in the different transits. This (somehow exotic) idea has already been investigated by \cite{zhu14} for candidates with short-cadence information.  Also, \cite{correia14} studied the effects of non-spheric close-in planets on the transit signal. They suggested that the detection of the non-spherical shape could reveal the internal structure of the planet. This scenario would require several strong constraints on the planet rotation period. However, short-cadence (1 minute) precise photometric data is needed to measure the relevance of this effect. \\

\item {\it Starspots \& spin-orbit orientation}.- The occultation of  { starspots} by the planet during its grazing transit increases the observed relative flux in the LC and could { cause the presence of brighter data points in the light curve. This scenario can be divided into two configurations:}

\begin{itemize}

\item {\it Spin-orbit alignment.-} In the case of spin-orbit alignment (where the orbit of the planet is perpendicular to the rotational axis of the star), the transit of the planet would only occult the higher latitudes of the star. This option would imply the existence of high latitude (polar) starspots remaining active during several months.  Depending on the impact parameter of the transit we could measure the lowest latitudes covered by the planet. For instance, in the case of Kepler-447 the lowest region of the stellar disk covered by the planet would be $b_{\rm min}=b-R_p/R_{\star}=0.911$, corresponding to a latitude of $\sim 84^{\circ}$. Polar starspots have been found in main sequence stars \citep[e.g., ][]{jeffers02} and stars with rapid rotation \citep{schuessler92}. In particular, \cite{sanchis-ojeda13} detected spot-crossing events in the transits of the hot Jupiter Kepler-63b. The almost perpendicular orientation of this orbit with respect to the spin axis of the star indicated the presence of a long-lived starspot at one of the star rotation poles. The existence and persistence of these polar spots in Solar analogs has been previously theorized based on simulations \citep[e.g., ][and references there in]{brown10} and have been detected in few cases using different techniques \citep{strassmeier98}. \\

\item {\it Spin-orbit misalignment.-} The misalignment of the stellar rotation axis and the orbital angular momentum has been observed in planetary systems \citep[e.g.,][]{sanchis-ojeda11} and particularly in hot-Jupiters \citep[e.g., ][]{hebrard08,triaud10}. A significant portion of these detections ($\sim 40$\%) resulted in obliquities larger than $\psi \sim 22.5^{\circ}$ \citep{addison14}. In such case, the grazing planetary transit can shade lower latitudes of the stellar disk, where stellar spots usually exist. The imprints of this effect on the radial velocity would be the presence of asymmetries in the Rossiter-McLaughlin effect \citep[see, e.g.,][]{queloz00}. If spot-crossing events are found in a planetary grazing transit this possibility can be tested by measuring the obliquity of the system with more accurate and higher-cadence RV measurements during the transit. However, we note that, as stated by \cite{winn10}, the great majority of planets around stars cooler than $T_{\rm eff}<6250$~K are well aligned.
\end{itemize}

\end{enumerate}

{ Due to the short duration of the transit ($T_{\rm dur}=1.135\pm0.016$ hours) and the long cadence observations for Kepler-447 (one data point every 29.4 minutes), it is not possible to perform an individual analysis of the different transits and thus test any of these interesting configurations. Short-cadence photometry would, instead, allow an individual fitting of the single transits.}

\subsection{The effect of cosmic rays in short-duration transits \label{sec:outliers}}

\begin{figure*}[htbf]
   \centering
   \includegraphics[width=0.5\textwidth]{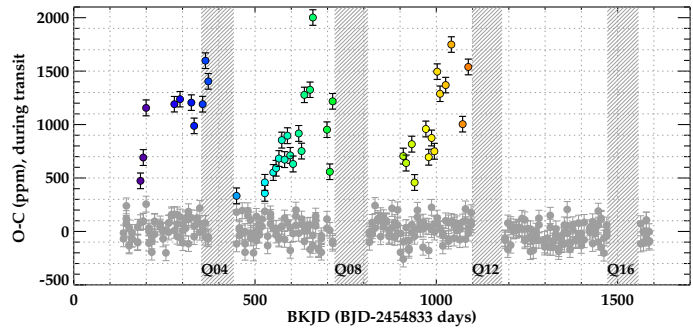}
   \caption{ Flux difference between the observed measurement and the transit model for data points inside the transit region. Outliers mostly due to the misidentification of cosmic rays by the \emph{Kepler} pipeline are marked in the same color code as in the right panel of Fig.~\ref{fig:rv}. The {\it Kepler} quarters of the mission with  few or no data for this KOI { due to the failure of CCD Module 3} are highlighted as gray shaded regions { (see Sect.~\S~\ref{sec:kepler})}.}
   \label{fig:LCdepths}
   \end{figure*}

{ The \emph{Kepler} pipeline processes the data and tries to correct for the presence of possible cosmic rays inside the aperture of a target. This correction is sometimes affected by erroneous identification of cosmic rays in the transit time interval of long-cadence data when the duration of the transit is short as compared to the cadence. When this situation happens, the pipeline tries to correct the photometry and erroneously assigns larger fluxes for the in-transit data in the mentioned configurations. The result of this in the phase-folded light curve is a transit somehow filled by outlier data points symmetrically distributed around the mid-transit time. These outliers in the transit region could be erroneously interpreted as hints for the detection of some of the configurations described in Sect.~\S~\ref{sec:configurations}. The \emph{Kepler} team provides information about which data points have been corrected for the possible existence of a cosmic ray by flagging them in the SAP\_QUALITY field of the delivered data files with the flag 128\footnote{See description of all SAP\_QUALITY flags in the latest version of the Kepler archive manual in https://archive.stsci.edu/kepler/manuals/archive\_manual.pdf}.

Due to the very specific situations in which this effect can play an important role and contaminate the transit region of a planetary transit, there are few cases where this instrumental effect has been reported. In particular, in \cite{herrero13}, the authors misidentify the erroneously corrected data points in the eclipse interval with a spot-crossing event in the system LHS 6343. In a recent work by \cite{montet14}, the authors identified the data points corresponding to this spot-crossing event in LHS 6343 as flagged data points with the SAP\_QUALITY equal to 128, and thus being cosmic rays misidentifications by the \emph{Kepler} pipeline. 

In the case of Kepler-447, the version of the pipeline used in the first twelve quarters produced this kind of misidentifications, while no erroneous detection was found in the remaining quarters. The $\sim$12\% of the data points inside the transit time interval along the whole \emph{Kepler} mission where tentatively identified (and thus corrected by the pipeline) as cosmic rays (see Fig.~\ref{fig:LCdepths}). The result, if the quality flags are not taken into account, is a transit somehow filled by outlier data points (see Fig.~\ref{fig:rv}, right panel) that could be erroneously interpreted as true signals caused by the configurations analyzed in Sect.~\S~\ref{sec:configurations}. Besides, owing to the extremely large impact parameter of the transit of Kepler-447b, thus being a planetary grazing transit, these flagged data points could by chance be mimicking the signal induced by some of those configurations. It is thus worthy to warn about this effect and to highlight the need of taking into account the quality flags provided by the \emph{Kepler} team in short-duration transit events. }

\section{Summary and conclusions \label{sec:conclusions}}

We have confirmed the planetary nature of Kepler-447b by analyzing its radial velocity imprint on its G8V host star and the periodic dips observed in the {\it Kepler} light curve. Our simultaneous modeling of both signals determined that Kepler-447b is a giant planet with a mass of $M_p = 1.35^{+0.48}_{-0.46}$ and a radius of $R_p=1.65^{+0.59}_{-0.56}$, orbiting in a slightly inclined orbit. This, combined to its closeness to the host star ($a/R_{\star}\sim 20.4$), implies an extremely grazing transit, with a grazing criterion of $b+R_p/R_{\star} > 1$, above the $3\sigma$ level. This is thus the most grazing transit known so far and the second confirmed planet to accomplish this criterion \citep[after WASP-67 b, ][]{mancini14}. Thus, we have found and characterized the first planetary grazing transit in the very large crop produced by {\it Kepler}.

{ The extremely large impact parameter found for this planet can help in the search for additional effects playing a role in the system. We have discussed four interesting configurations that could be detectable by obtaining high-accurate and short-cadence photometry of grazing planetary transits. These include additional perturbing bodies (such as outer planets, exomoons, Trojan bodies, etc.), stellar pulsations changing the stellar radius and thus the impact parameter, planet rotation and oblateness, or spot crossing events. The non-negligible eccentricity found for this close-in giant planet suggests the presence of additional companion/s in the system. Thus, future works obtaining short cadence photometry with the typical \emph{Kepler} accuracy and during a relatively long time span would allow individual transit analysis and look for transit depth and duration variations { as well as possible spot crossing events}.}


\begin{acknowledgements}
      This research has been funded by Spanish grant  AYA2012-38897-C02-01. J. Lillo-Box thanks the CSIC JAE-predoc programme for the PhD fellowship. We also thank Calar Alto Observatory, both the open TAC and Spanish GTO panel, for allocating our observing runs, and the CAHA staff for their effort and passion in their work, being very helpful during our visitor and service CAFE observations. J. Lillo-Box thanks H. Bouy for its setting and handling of the CAB's computer cluster. PF and NCS acknowledge support by  Funda\c{c}\~ao para a Ci\^encia e a Tecnologia (FCT) through Investigador FCT contracts of reference IF/01037/2013 and IF/00169/2012, respectively, and POPH/FSE (EC) by FEDER funding through the program ``Programa Operacional de Factores de Competitividade - COMPETE''. We also acknowledge the support from the European Research Council/European Community under the FP7 through Starting Grant agreement number 239953. This work used the python packages {\it numpy}, {\it astroML}, and {\it asciitable}. We do appreciate the extraordinary effort and generosity by all people involved in the {\it Kepler} mission at some point, both by running the observatory and  by providing an amazing database with a hundred of thousands  light curves and thousands of planetary candidates, so the community can work on them, and confirm and  characterize the largest and more comprehensive planetary sample found so far. { We also thank the referee for his/her useful comments and suggestions that have improved the quality of this paper. We would like to thank the kind comments and suggestions by A. Vanderburg, D. Ragozzine, E. Kruse, and J. Southworth on the analysis of the misidentified cosmic rays by the \emph{Kepler} pipeline.}
\end{acknowledgements}

\bibliographystyle{aa} 
\bibliography{biblio2} 

\begin{thebibliography}{58}
\expandafter\ifx\csname natexlab\endcsname\relax\def\natexlab#1{#1}\fi

\bibitem[{{Aceituno} {et~al.}(2013){Aceituno}, {S{\'a}nchez}, {Grupp}, {Lillo},
  {Hern{\'a}n-Obispo}, {Benitez}, {Montoya}, {Thiele}, {Pedraz}, {Barrado},
  {Dreizler}, \& {Bean}}]{aceituno13}
{Aceituno}, J., {S{\'a}nchez}, S.~F., {Grupp}, F., {et~al.} 2013, \aap, 552,
  A31

\bibitem[{{Addison} {et~al.}(2014){Addison}, {Tinney}, {Wright}, \&
  {Bayliss}}]{addison14}
{Addison}, B.~C., {Tinney}, C.~G., {Wright}, D.~J., \& {Bayliss}, D. 2014,
  \apj, 792, 112

\bibitem[{{Anderson} {et~al.}(2011){Anderson}, {Barros}, {Boisse}, {Bouchy},
  {Collier Cameron}, {Faedi}, {Hebrard}, {Hellier}, {Lendl}, {Moutou},
  {Pollacco}, {Santerne}, {Smalley}, {Smith}, {Todd}, {Triaud}, {West},
  {Wheatley}, {Bento}, {Enoch}, {Gillon}, {Maxted}, {McCormac}, {Queloz},
  {Simpson}, \& {Skillen}}]{anderson11}
{Anderson}, D.~R., {Barros}, S.~C.~C., {Boisse}, I., {et~al.} 2011, \pasp, 123,
  555

\bibitem[{{Ballard} {et~al.}(2014){Ballard}, {Chaplin}, {Charbonneau},
  {D{\'e}sert}, {Fressin}, {Zeng}, {Werner}, {Davies}, {Silva Aguirre}, {Basu},
  {Christensen-Dalsgaard}, {Metcalfe}, {Stello}, {Bedding}, {Campante},
  {Handberg}, {Karoff}, {Elsworth}, {Gilliland}, {Hekker}, {Huber}, {Kawaler},
  {Kjeldsen}, {Lund}, \& {Lundkvist}}]{ballard14}
{Ballard}, S., {Chaplin}, W.~J., {Charbonneau}, D., {et~al.} 2014, \apj, 790,
  12

\bibitem[{{Baranne} {et~al.}(1996){Baranne}, {Queloz}, {Mayor}, {Adrianzyk},
  {Knispel}, {Kohler}, {Lacroix}, {Meunier}, {Rimbaud}, \& {Vin}}]{baranne96}
{Baranne}, A., {Queloz}, D., {Mayor}, M., {et~al.} 1996, \aaps, 119, 373

\bibitem[{{Barclay} {et~al.}(2013){Barclay}, {Rowe}, {Lissauer}, {Huber},
  {Fressin}, {Howell}, {Bryson}, {Chaplin}, {D{\'e}sert}, {Lopez}, {Marcy},
  {Mullally}, {Ragozzine}, {Torres}, {Adams}, {Agol}, {Barrado}, {Basu},
  {Bedding}, {Buchhave}, {Charbonneau}, {Christiansen},
  {Christensen-Dalsgaard}, {Ciardi}, {Cochran}, {Dupree}, {Elsworth},
  {Everett}, {Fischer}, {Ford}, {Fortney}, {Geary}, {Haas}, {Handberg},
  {Hekker}, {Henze}, {Horch}, {Howard}, {Hunter}, {Isaacson}, {Jenkins},
  {Karoff}, {Kawaler}, {Kjeldsen}, {Klaus}, {Latham}, {Li}, {Lillo-Box},
  {Lund}, {Lundkvist}, {Metcalfe}, {Miglio}, {Morris}, {Quintana}, {Stello},
  {Smith}, {Still}, \& {Thompson}}]{barclay13}
{Barclay}, T., {Rowe}, J.~F., {Lissauer}, J.~J., {et~al.} 2013, \nat, 494, 452

\bibitem[{{B{\'e}ky} {et~al.}(2011){B{\'e}ky}, {Bakos}, {Hartman}, {Torres},
  {Latham}, {Jord{\'a}n}, {Arriagada}, {Bayliss}, {Kiss}, {Kov{\'a}cs},
  {Quinn}, {Marcy}, {Howard}, {Fischer}, {Johnson}, {Esquerdo}, {Noyes},
  {Buchhave}, {Sasselov}, {Stefanik}, {Perumpilly}, {L{\'a}z{\'a}r}, {Papp}, \&
  {S{\'a}ri}}]{beky11}
{B{\'e}ky}, B., {Bakos}, G.~{\'A}., {Hartman}, J., {et~al.} 2011, \apj, 734,
  109

\bibitem[{{Borucki} {et~al.}(2012){Borucki}, {Koch}, {Batalha}, {Bryson},
  {Rowe}, {Fressin}, {Torres}, {Caldwell}, {Christensen-Dalsgaard}, {Cochran},
  {DeVore}, {Gautier}, {Geary}, {Gilliland}, {Gould}, {Howell}, {Jenkins},
  {Latham}, {Lissauer}, {Marcy}, {Sasselov}, {Boss}, {Charbonneau}, {Ciardi},
  {Kaltenegger}, {Doyle}, {Dupree}, {Ford}, {Fortney}, {Holman}, {Steffen},
  {Mullally}, {Still}, {Tarter}, {Ballard}, {Buchhave}, {Carter},
  {Christiansen}, {Demory}, {D{\'e}sert}, {Dressing}, {Endl}, {Fabrycky},
  {Fischer}, {Haas}, {Henze}, {Horch}, {Howard}, {Isaacson}, {Kjeldsen},
  {Johnson}, {Klaus}, {Kolodziejczak}, {Barclay}, {Li}, {Meibom}, {Prsa},
  {Quinn}, {Quintana}, {Robertson}, {Sherry}, {Shporer}, {Tenenbaum},
  {Thompson}, {Twicken}, {Van Cleve}, {Welsh}, {Basu}, {Chaplin}, {Miglio},
  {Kawaler}, {Arentoft}, {Stello}, {Metcalfe}, {Verner}, {Karoff}, {Lundkvist},
  {Lund}, {Handberg}, {Elsworth}, {Hekker}, {Huber}, {Bedding}, \&
  {Rapin}}]{borucki12}
{Borucki}, W.~J., {Koch}, D.~G., {Batalha}, N., {et~al.} 2012, \apj, 745, 120

\bibitem[{{Brown} {et~al.}(2010){Brown}, {Browning}, {Brun}, {Miesch}, \&
  {Toomre}}]{brown10}
{Brown}, B.~P., {Browning}, M.~K., {Brun}, A.~S., {Miesch}, M.~S., \& {Toomre},
  J. 2010, \apj, 711, 424

\bibitem[{{Burke} {et~al.}(2014){Burke}, {Bryson}, {Mullally}, {Rowe},
  {Christiansen}, {Thompson}, {Coughlin}, {Haas}, {Batalha}, {Caldwell},
  {Jenkins}, {Still}, {Barclay}, {Borucki}, {Chaplin}, {Ciardi}, {Clarke},
  {Cochran}, {Demory}, {Esquerdo}, {Gautier}, {Gilliland}, {Girouard}, {Havel},
  {Henze}, {Howell}, {Huber}, {Latham}, {Li}, {Morehead}, {Morton}, {Pepper},
  {Quintana}, {Ragozzine}, {Seader}, {Shah}, {Shporer}, {Tenenbaum}, {Twicken},
  \& {Wolfgang}}]{burke14}
{Burke}, C.~J., {Bryson}, S.~T., {Mullally}, F., {et~al.} 2014, \apjs, 210, 19

\bibitem[{{Ciceri} {et~al.}(2015){Ciceri}, {Lillo-Box}, {Southworth},
  {Mancini}, {Henning}, \& {Barrado}}]{ciceri14}
{Ciceri}, S., {Lillo-Box}, J., {Southworth}, J., {et~al.} 2015, \aap, 573, L5

\bibitem[{{Claret} \& {Bloemen}(2011)}]{claret11}
{Claret}, A. \& {Bloemen}, S. 2011, \aap, 529, A75

\bibitem[{{Correia}(2014)}]{correia14}
{Correia}, A.~C.~M. 2014, \aap, 570, L5

\bibitem[{{Cruz} {et~al.}(2015){Cruz}, {Barrado}, {Lillo-Box}, {Diaz},
  {Birkby}, {L{\'o}pez-Morales}, {Hodgkin}, \& {Fortney}}]{cruz14}
{Cruz}, P., {Barrado}, D., {Lillo-Box}, J., {et~al.} 2015, \aap, 574, A103

\bibitem[{{Delbouille}(1971)}]{delbouille72}
{Delbouille}, L. 1971, Transactions of the International Astronomical Union,
  Series B, 14, 116

\bibitem[{{D{\'e}sert} {et~al.}(2011){D{\'e}sert}, {Charbonneau}, {Demory},
  {Ballard}, {Carter}, {Fortney}, {Cochran}, {Endl}, {Quinn}, {Isaacson},
  {Fressin}, {Buchhave}, {Latham}, {Knutson}, {Bryson}, {Torres}, {Rowe},
  {Batalha}, {Borucki}, {Brown}, {Caldwell}, {Christiansen}, {Deming},
  {Fabrycky}, {Ford}, {Gilliland}, {Gillon}, {Haas}, {Jenkins}, {Kinemuchi},
  {Koch}, {Lissauer}, {Lucas}, {Mullally}, {MacQueen}, {Marcy}, {Sasselov},
  {Seager}, {Still}, {Tenenbaum}, {Uddin}, \& {Winn}}]{desert11b}
{D{\'e}sert}, J.-M., {Charbonneau}, D., {Demory}, B.-O., {et~al.} 2011, \apjs,
  197, 14

\bibitem[{{Figueira} {et~al.}(2013){Figueira}, {Santos}, {Pepe}, {Lovis}, \&
  {Nardetto}}]{figueira13}
{Figueira}, P., {Santos}, N.~C., {Pepe}, F., {Lovis}, C., \& {Nardetto}, N.
  2013, \aap, 557, A93

\bibitem[{{H{\'e}brard} {et~al.}(2008){H{\'e}brard}, {Bouchy}, {Pont},
  {Loeillet}, {Rabus}, {Bonfils}, {Moutou}, {Boisse}, {Delfosse}, {Desort},
  {Eggenberger}, {Ehrenreich}, {Forveille}, {Lagrange}, {Lovis}, {Mayor},
  {Pepe}, {Perrier}, {Queloz}, {Santos}, {S{\'e}gransan}, {Udry}, \&
  {Vidal-Madjar}}]{hebrard08}
{H{\'e}brard}, G., {Bouchy}, F., {Pont}, F., {et~al.} 2008, \aap, 488, 763

\bibitem[{{Hellier} {et~al.}(2012){Hellier}, {Anderson}, {Collier Cameron},
  {Doyle}, {Fumel}, {Gillon}, {Jehin}, {Lendl}, {Maxted}, {Pepe}, {Pollacco},
  {Queloz}, {S{\'e}gransan}, {Smalley}, {Smith}, {Southworth}, {Triaud},
  {Udry}, \& {West}}]{hellier12}
{Hellier}, C., {Anderson}, D.~R., {Collier Cameron}, A., {et~al.} 2012, \mnras,
  426, 739

\bibitem[{{Herrero} {et~al.}(2013){Herrero}, {Lanza}, {Ribas}, {Jordi}, \&
  {Morales}}]{herrero13}
{Herrero}, E., {Lanza}, A.~F., {Ribas}, I., {Jordi}, C., \& {Morales}, J.~C.
  2013, \aap, 553, A66

\bibitem[{{Huber} {et~al.}(2014){Huber}, {Silva Aguirre}, {Matthews},
  {Pinsonneault}, {Gaidos}, {Garc{\'{\i}}a}, {Hekker}, {Mathur}, {Mosser},
  {Torres}, {Bastien}, {Basu}, {Bedding}, {Chaplin}, {Demory}, {Fleming},
  {Guo}, {Mann}, {Rowe}, {Serenelli}, {Smith}, \& {Stello}}]{huber14}
{Huber}, D., {Silva Aguirre}, V., {Matthews}, J.~M., {et~al.} 2014, \apjs, 211,
  2

\bibitem[{{Jackson} {et~al.}(2008){Jackson}, {Greenberg}, \&
  {Barnes}}]{jackson08}
{Jackson}, B., {Greenberg}, R., \& {Barnes}, R. 2008, \apj, 681, 1631

\bibitem[{{Jeffers} {et~al.}(2002){Jeffers}, {Barnes}, \& {Collier
  Cameron}}]{jeffers02}
{Jeffers}, S.~V., {Barnes}, J.~R., \& {Collier Cameron}, A. 2002, \mnras, 331,
  666

\bibitem[{{Kipping}(2009)}]{kipping09b}
{Kipping}, D.~M. 2009, \mnras, 396, 1797

\bibitem[{{Kipping}(2010)}]{kipping10}
{Kipping}, D.~M. 2010, \mnras, 408, 1758

\bibitem[{{Laughlin} \& {Chambers}(2002)}]{laughlin02}
{Laughlin}, G. \& {Chambers}, J.~E. 2002, Astronomical Journal, 124, 592

\bibitem[{{Lillo-Box} {et~al.}(2012){Lillo-Box}, {Barrado}, \&
  {Bouy}}]{lillo-box12}
{Lillo-Box}, J., {Barrado}, D., \& {Bouy}, H. 2012, \aap, 546, A10

\bibitem[{{Lillo-Box} {et~al.}(2014{\natexlab{a}}){Lillo-Box}, {Barrado}, \&
  {Bouy}}]{lillo-box14b}
{Lillo-Box}, J., {Barrado}, D., \& {Bouy}, H. 2014{\natexlab{a}}, \aap, 566,
  A103

\bibitem[{{Lillo-Box} {et~al.}(2014{\natexlab{b}}){Lillo-Box}, {Barrado},
  {Henning}, {Mancini}, {Ciceri}, {Figueira}, {Santos}, {Aceituno}, \&
  {S{\'a}nchez}}]{lillo-box14c}
{Lillo-Box}, J., {Barrado}, D., {Henning}, T., {et~al.} 2014{\natexlab{b}},
  \aap, 568, L1

\bibitem[{{Lillo-Box} {et~al.}(2015){Lillo-Box}, {Barrado}, {Mancini},
  {Henning}, {Figueira}, {Ciceri}, \& {Santos}}]{lillo-box15a}
{Lillo-Box}, J., {Barrado}, D., {Mancini}, L., {et~al.} 2015, ArXiv e-prints,
  1501.05183

\bibitem[{{Lillo-Box} {et~al.}(2014{\natexlab{c}}){Lillo-Box}, {Barrado},
  {Moya}, {Montesinos}, {Montalb{\'a}n}, {Bayo}, {Barbieri}, {R{\'e}gulo},
  {Mancini}, {Bouy}, \& {Henning}}]{lillo-box14}
{Lillo-Box}, J., {Barrado}, D., {Moya}, A., {et~al.} 2014{\natexlab{c}}, \aap,
  562, A109

\bibitem[{{Lissauer} {et~al.}(2011){Lissauer}, {Fabrycky}, {Ford}, {Borucki},
  {Fressin}, {Marcy}, {Orosz}, {Rowe}, {Torres}, {Welsh}, {Batalha}, {Bryson},
  {Buchhave}, {Caldwell}, {Carter}, {Charbonneau}, {Christiansen}, {Cochran},
  {Desert}, {Dunham}, {Fanelli}, {Fortney}, {Gautier}, {Geary}, {Gilliland},
  {Haas}, {Hall}, {Holman}, {Koch}, {Latham}, {Lopez}, {McCauliff}, {Miller},
  {Morehead}, {Quintana}, {Ragozzine}, {Sasselov}, {Short}, \&
  {Steffen}}]{lissauer11}
{Lissauer}, J.~J., {Fabrycky}, D.~C., {Ford}, E.~B., {et~al.} 2011, \nat, 470,
  53

\bibitem[{{Mancini} {et~al.}(2014){Mancini}, {Southworth}, {Ciceri}, {Calchi
  Novati}, {Dominik}, {Henning}, {J{\o}rgensen}, {Korhonen}, {Nikolov},
  {Alsubai}, {Bozza}, {Bramich}, {D'Ago}, {Figuera Jaimes}, {Galianni}, {Gu},
  {Harps{\o}e}, {Hinse}, {Hundertmark}, {Juncher}, {Kains}, {Popovas}, {Rabus},
  {Rahvar}, {Skottfelt}, {Snodgrass}, {Street}, {Surdej}, {Tsapras}, {Vilela},
  {Wang}, \& {Wertz}}]{mancini14}
{Mancini}, L., {Southworth}, J., {Ciceri}, S., {et~al.} 2014, \aap, 568, A127

\bibitem[{{Mandel} \& {Agol}(2002)}]{mandel02}
{Mandel}, K. \& {Agol}, E. 2002, \apjl, 580, L171

\bibitem[{{McQuillan} {et~al.}(2013{\natexlab{a}}){McQuillan}, {Aigrain}, \&
  {Mazeh}}]{mcquillan13a}
{McQuillan}, A., {Aigrain}, S., \& {Mazeh}, T. 2013{\natexlab{a}}, \mnras, 432,
  1203

\bibitem[{{McQuillan} {et~al.}(2013{\natexlab{b}}){McQuillan}, {Mazeh}, \&
  {Aigrain}}]{mcquillan13b}
{McQuillan}, A., {Mazeh}, T., \& {Aigrain}, S. 2013{\natexlab{b}}, \apjl, 775,
  L11

\bibitem[{{Montet} {et~al.}(2014){Montet}, {Johnson}, {Muirhead}, {Villar},
  {Vassallo}, {Baranec}, {Law}, {Riddle}, {Marcy}, {Howard}, \&
  {Isaacson}}]{montet14}
{Montet}, B.~T., {Johnson}, J.~A., {Muirhead}, P.~S., {et~al.} 2014, ArXiv
  e-prints

\bibitem[{{M{\"u}ller} {et~al.}(2013){M{\"u}ller}, {Huber}, {Czesla}, {Wolter},
  \& {Schmitt}}]{muller13}
{M{\"u}ller}, H.~M., {Huber}, K.~F., {Czesla}, S., {Wolter}, U., \& {Schmitt},
  J.~H.~M.~M. 2013, \aap, 560, A112

\bibitem[{{Ortiz} {et~al.}(2015){Ortiz}, {Gandolfi}, {Reffert}, {Quirrenbach},
  {Deeg}, {Karjalainen}, {Monta{\~n}{\'e}s-Rodr{\'{\i}}guez}, {Nespral},
  {Nowak}, {Osorio}, \& {Palle}}]{ortiz14}
{Ortiz}, M., {Gandolfi}, D., {Reffert}, S., {et~al.} 2015, \aap, 573, L6

\bibitem[{{Pepe} {et~al.}(2011){Pepe}, {Lovis}, {S{\'e}gransan}, {Benz},
  {Bouchy}, {Dumusque}, {Mayor}, {Queloz}, {Santos}, \& {Udry}}]{pepe11}
{Pepe}, F., {Lovis}, C., {S{\'e}gransan}, D., {et~al.} 2011, \aap, 534, A58

\bibitem[{{Queloz} {et~al.}(2001){Queloz}, {Henry}, {Sivan}, {Baliunas},
  {Beuzit}, {Donahue}, {Mayor}, {Naef}, {Perrier}, \& {Udry}}]{queloz01}
{Queloz}, D., {Henry}, G.~W., {Sivan}, J.~P., {et~al.} 2001, \aap, 379, 279

\bibitem[{{Queloz} {et~al.}(2000){Queloz}, {Mayor}, {Weber}, {Bl{\'e}cha},
  {Burnet}, {Confino}, {Naef}, {Pepe}, {Santos}, \& {Udry}}]{queloz00}
{Queloz}, D., {Mayor}, M., {Weber}, L., {et~al.} 2000, \aap, 354, 99

\bibitem[{{Riviere-Marichalar} {et~al.}(2014){Riviere-Marichalar}, {Barrado},
  {Montesinos}, {Duch{\^e}ne}, {Bouy}, {Pinte}, {Menard}, {Donaldson}, {Eiroa},
  {Krivov}, {Kamp}, {Mendigut{\'{\i}}a}, {Dent}, \&
  {Lillo-Box}}]{riviere-marichalar14}
{Riviere-Marichalar}, P., {Barrado}, D., {Montesinos}, B., {et~al.} 2014, \aap,
  565, A68

\bibitem[{{Riviere-Marichalar} {et~al.}(2013){Riviere-Marichalar}, {Pinte},
  {Barrado}, {Thi}, {Eiroa}, {Kamp}, {Montesinos}, {Donaldson}, {Augereau},
  {Hu{\'e}lamo}, {Roberge}, {Ardila}, {Sandell}, {Williams}, {Dent}, {Menard},
  {Lillo-Box}, \& {Duch{\^e}ne}}]{riviere-marichalar13}
{Riviere-Marichalar}, P., {Pinte}, C., {Barrado}, D., {et~al.} 2013, \aap, 555,
  A67

\bibitem[{{Sanchis-Ojeda} \& {Winn}(2011)}]{sanchis-ojeda11}
{Sanchis-Ojeda}, R. \& {Winn}, J.~N. 2011, \apj, 743, 61

\bibitem[{{Sanchis-Ojeda} {et~al.}(2013){Sanchis-Ojeda}, {Winn}, {Marcy},
  {Howard}, {Isaacson}, {Johnson}, {Torres}, {Albrecht}, {Campante}, {Chaplin},
  {Davies}, {Lund}, {Carter}, {Dawson}, {Buchhave}, {Everett}, {Fischer},
  {Geary}, {Gilliland}, {Horch}, {Howell}, \& {Latham}}]{sanchis-ojeda13}
{Sanchis-Ojeda}, R., {Winn}, J.~N., {Marcy}, G.~W., {et~al.} 2013, \apj, 775,
  54

\bibitem[{{Santos} {et~al.}(2014){Santos}, {Mortier}, {Faria}, {Dumusque},
  {Adibekyan}, {Delgado-Mena}, {Figueira}, {Benamati}, {Boisse}, {Cunha},
  {Gomes da Silva}, {Lo Curto}, {Lovis}, {Martins}, {Mayor}, {Melo}, {Oshagh},
  {Pepe}, {Queloz}, {Santerne}, {S{\'e}gransan}, {Sozzetti}, {Sousa}, \&
  {Udry}}]{santos14}
{Santos}, N.~C., {Mortier}, A., {Faria}, J.~P., {et~al.} 2014, \aap, 566, A35

\bibitem[{{Schuessler} \& {Solanki}(1992)}]{schuessler92}
{Schuessler}, M. \& {Solanki}, S.~K. 1992, \aap, 264, L13

\bibitem[{{Silva}(2003)}]{silva03}
{Silva}, A.~V.~R. 2003, \apjl, 585, L147

\bibitem[{{Silva-Valio} \& {Lanza}(2011)}]{silva-valio11}
{Silva-Valio}, A. \& {Lanza}, A.~F. 2011, \aap, 529, A36

\bibitem[{{Smalley} {et~al.}(2011){Smalley}, {Anderson}, {Collier Cameron},
  {Hellier}, {Lendl}, {Maxted}, {Queloz}, {Triaud}, {West}, {Bentley}, {Enoch},
  {Gillon}, {Lister}, {Pepe}, {Pollacco}, {Segransan}, {Smith}, {Southworth},
  {Udry}, {Wheatley}, {Wood}, \& {Bento}}]{smalley11}
{Smalley}, B., {Anderson}, D.~R., {Collier Cameron}, A., {et~al.} 2011, \aap,
  526, A130

\bibitem[{{Strassmeier} \& {Rice}(1998)}]{strassmeier98}
{Strassmeier}, K.~G. \& {Rice}, J.~B. 1998, \aap, 330, 685

\bibitem[{{Tassoul}(1987)}]{tassoul87}
{Tassoul}, J.-L. 1987, \apj, 322, 856

\bibitem[{{Torres}(2010)}]{torres10}
{Torres}, G. 2010, \aj, 140, 1158

\bibitem[{{Triaud} {et~al.}(2010){Triaud}, {Collier Cameron}, {Queloz},
  {Anderson}, {Gillon}, {Hebb}, {Hellier}, {Loeillet}, {Maxted}, {Mayor},
  {Pepe}, {Pollacco}, {S{\'e}gransan}, {Smalley}, {Udry}, {West}, \&
  {Wheatley}}]{triaud10}
{Triaud}, A.~H.~M.~J., {Collier Cameron}, A., {Queloz}, D., {et~al.} 2010,
  \aap, 524, A25

\bibitem[{{Winn} {et~al.}(2010){Winn}, {Fabrycky}, {Albrecht}, \&
  {Johnson}}]{winn10}
{Winn}, J.~N., {Fabrycky}, D., {Albrecht}, S., \& {Johnson}, J.~A. 2010, \apjl,
  718, L145

\bibitem[{{Zahn}(1966)}]{zahn66}
{Zahn}, J.~P. 1966, Annales d'Astrophysique, 29, 565

\bibitem[{{Zhu} {et~al.}(2014){Zhu}, {Huang}, {Zhou}, \& {Lin}}]{zhu14}
{Zhu}, W., {Huang}, C.~X., {Zhou}, G., \& {Lin}, D.~N.~C. 2014, \apj, 796, 67

\end{thebibliography}

\begin{table}[HT]
\setlength{\extrarowheight}{7pt}
\tiny
\caption{ Stellar parameters derived by different techniques. \label{tab:stellarparams} }
\centering
\begin{tabular}{rcc}
\hline\hline

Parameter                 & Huber2014                   & This work (\S~\ref{sec:stellarparams})      \\ \hline
$T_{\rm eff}$ (K)         & $5555^{+171}_{-133}$         & $5493\pm62$       \\
$log(g)$ (cgs)            & $4.440^{+0.119}_{-0.308}$    & $4.40\pm0.10$      \\ 
$[Fe/H]$                   & $-0.40^{+0.36}_{-0.26}$      & $0.07\pm0.05$      \\                          
$M_{\star}$ ($M_{\odot}$) & $0.764^{+0.145}_{-0.049}$       & $1.00\pm0.21$                \\       
$R_{\star}$ ($R_{\odot}$) & $0.872^{+0.419}_{-0.120}$       & $1.03\pm0.16$                 \\       

\hline
\hline

\end{tabular}
\end{table}

\begin{table}[h]
\setlength{\extrarowheight}{7pt}
\small
\caption{ Parameters of the joint fitting of the radial velocity and transit data. \label{tab:results} }
\centering
\begin{tabular}{rccc}
\hline\hline

Parameter        & Range          & Value\tablefootmark{a} & Units     \\  \hline
$K$              & $[50,250]$     & $ 141^{+42}_{-42} $           &  m/s           \\
$e$              & $[0.0,0.5]$    & $ 0.123^{+0.037}_{-0.036} $   &  \\
$\omega$         & $[0,360]$      & $ 98.3^{+1.1}_{-11.0} $       & deg.  \\
$V_{sys}$        & $[1.0,2.0]$    & $ 1.39^{+0.11}_{-0.11} $       & km/s    \\ 
$a/R_{\star}$    & $[10,40]$      & $ 20.41^{+0.36}_{-0.19} $      &   \\
$v_{\rm off1}$   & $[-1000,1000]$ & $ -363^{+88}_{-20} $     & m/s  \\
$v_{\rm off2}$   & $[-1000,1000]$ & $ -84^{+24}_{-25} $            & m/s  \\
$R_p/R_{\star}$  & $[0.01,0.50]$  & $ 0.165^{+0.049}_{-0.048} $    &  \\
$i$              & $[80,90]$     &  $ 86.55^{+0.24}_{-0.32} $    & deg.  \\
$\phi_{\rm off}$ & $[10,20]$      & $ 1.22^{+0.37}_{-0.11} $       & min.  \\                             

\hline
\hline

\end{tabular}
\tablefoot{
\tablefoottext{a}{Uncertainties are $3\sigma$.}
}
\end{table}

 \begin{table}[h]
\setlength{\extrarowheight}{7pt}
\small
\caption{ Derived physical and orbital parameters from the fitted solution of the joint analysis.  \label{tab:derived} }
\centering
\begin{tabular}{rcccl}
\hline\hline

Parameter      &   \multicolumn{2}{c}{Derived parameters} & Units            & Assumptions\tablefootmark{(c)}     \\  
      &   \textbf{Sol. A}\tablefootmark{(a)}  & Sol. B\tablefootmark{(b)}  &             &    \\  \hline

$R_{\star}$    &        $ \mathbf{1.05^{+0.19}_{-0.19}} $  &       $ 0.915^{+0.143}_{-0.100} $         & $R_{\odot}$        &  $M_{\star}$, $\log{g}$     \\
$a$            &   $ \mathbf{0.0769^{+0.0062}_{-0.0079}} $ &    $ 0.0703^{+0.0036}_{-0.0023} $         & AU                & $P_{\rm orb}$, $M_{\star}$    \\
$b$            &         $ \mathbf{1.076^{+0.112}_{-0.086}} $ &          $ 1.076^{+0.112}_{-0.086} $        &                   & $a/R_{\star}$, $i$, $e$, $\omega$     \\ 
$T_{\rm dur}$    &  $ \mathbf{1.135^{+0.016}_{-0.016}} $    &       $ 1.135^{+0.016}_{-0.016} $          & hr                & TR model \\
$M_p$          &         $ \mathbf{1.37^{+0.48}_{-0.46}} $ &          $ 1.14^{+0.38}_{-0.34} $        & $M_{\rm Jup}$       & $K$, $M_{\star}$, $P_{\rm orb}$       \\
$R_p$          &         $ \mathbf{1.65^{+0.59}_{-0.56}} $ &          $ 1.40^{+0.89}_{-0.42} $        & $R_{\rm Jup}$       & $R_p/R_{\star}$, $R_{\star}$       \\
<$\rho_p$>     &         $ \mathbf{0.30^{+0.71}_{-0.24}} $ &          $ 0.42^{+0.74}_{-0.37} $        & $\rho_{\rm Jup}$    & $M_p$, $R_p$      \\
$A_{\rm ecl}/A_{\rm pl}^{\dagger}$      &  $ \mathbf{20^{+30}_{-23}} $ &                $ 20^{+30}_{-23} $        & \%                & $b$, $R_p/R_{\star}$ \\                                        
\hline
\hline

\end{tabular}
\tablefoot{All uncertainties cover the 99.7\% of the probability distribution (i.e., $3\sigma$).
\tablefoottext{a}{Stellar radius and mass obtained from our spectroscopic analysis (see Sect.~\S~\ref{sec:stellarparams}).}
\tablefoottext{b}{Stellar radius and mass obtained from \cite{huber14}.}
\tablefoottext{c}{Assumed parameters in the estimation of the derived properties.}
{$\dagger$ Fraction of the planet's projected area eclipsing the stellar disk at mid-transit (in \%).}
}
\end{table}


\clearpage

   \clearpage

\end{document}